\documentclass[12pt]{article}

\newcommand{\zpr}{\mbox{$Z^{\prime}$}}
\newcommand{\upr}{\mbox{$U(1)^{\prime}$}}

\newcommand{\mz}{\mbox{$M_Z$}}

\newcommand{\x}{\mbox{$\times$}}

\begin{document}
\begin{flushright}
\baselineskip=12pt
UPR-1010-T \\
\end{flushright}

\begin{center}
\vglue 1.5cm

{\Large\bf Neutrino Masses in Supersymmetric $SU(3)_C\times
SU(2)_L \times U(1)_Y \times U(1)^{\prime}$ Models} \vglue 2.0cm
{\Large Junhai Kang$^a$, Paul Langacker$^a$ and Tianjun Li$^{a,
b}$} \vglue 1cm { $^a$ Department of Physics and Astronomy,
University of Pennsylvania, \\Philadelphia, PA 19104-6396, USA \\

$^b$ School of Natural Science, Institute for Advanced Study,  \\
             Einstein Drive, Princeton, NJ 08540, USA\\
}
\end{center}

\vglue 1.0cm
\begin{abstract}

We consider various possibilities for generating  neutrino masses
in supersymmetric models with an additional $U(1)^{\prime}$ gauge
symmetry. One class of  models involves two extra 
$U(1)^{\prime} \times U(1)''$ gauge symmetries,
with $U(1)''$ breaking at an intermediate scale and yielding small
Dirac masses through  high-dimensional operators. The right-handed
neutrinos $N^c_i$ can naturally decouple from the low energy $U(1)'$,
avoiding cosmological constraints.  A variant version can generate
large Majorana masses for $N^c_i$ and an ordinary see-saw. We
secondly consider models with a pair of heavy triplets which
couple to left-handed neutrinos. After integrating out the heavy
triplets, a small neutrino Majorana mass matrix can be generated
by the induced non-renormalizable terms. We also study models
involving the double-see-saw mechanism, in which heavy Majorana
masses for $N^c_i$ 
are associated with the TeV-scale of $U(1)'$ breaking. We
give the conditions to avoid runaway directions in such models and
discuss simple patterns for neutrino masses.

\end{abstract}

\vspace{0.5cm}
\begin{flushleft}
\baselineskip=12pt
\today\\
\end{flushleft}
\newpage
\baselineskip=14pt

\section{Introduction}

The possibility of an extra \upr \ gauge symmetry 
is well-motivated in superstring constructions~\cite{string},
grand unified theories~\cite{review}, models of dynamical symmetry
breaking~\cite{DSB}, little Higgs models~\cite{lHiggs}, and large extra
dimensions~\cite{led}. In supersymmetric models, an extra \upr \ can provide
an elegant solution to the $\mu$ problem~\cite{muprob1,muprob2},
with an effective $\mu$ parameter generated by the vacuum
expectation value (VEV) of the Standard Model (SM) singlet field
$S$ which breaks the \upr \ symmetry. This is somewhat similar to
the effective $\mu$ parameter in the Next to Minimal
Supersymmetric Standard Model (NMSSM)~\cite{NMSSM}. However, with
a \upr \ the extra discrete symmetries and their associated
cosmological domain wall problems~\cite{domainwall} 
associated with the NMSSM are absent\footnote{For other solutions,
see~\cite{nmssm}.}. A closely related feature is that the 
 Minimal Supersymmetric Standard Model
(MSSM) upper bound of \mz \ on
the tree-level mass of the corresponding lightest MSSM Higgs
scalar is relaxed, both in models with a \upr \ and in the NMSSM,
because of the Yukawa term $h S H_1 H_2$ in the superpotential~\cite{NMSSM} and
the $U(1)'$ $D$-term~\cite{Higgsbound}. More generally, for
specific \upr \ charge assignments for the ordinary and exotic
fields one can simultaneously ensure the absence of anomalies;
that all fields of the TeV-scale effective theory are chiral,
avoiding a generalized $\mu$ problem; and the absence of
dimension-4 proton decay operators~\cite{general}. $U(1)'$ models
can also be consistent with gauge unification and may have
implications in electroweak baryogenesis~\cite{BG}, cold dark matter~\cite{cdm}, rare
$B$ decays~\cite{rare}, and non-standard Higgs potentials~\cite{Higgs}.

There are stringent limits from direct searches at the
Tevatron~\cite{explim} and from indirect precision tests at the
$Z$-pole, at LEP 2, and from weak neutral current
experiments~\cite{indirect}. The constraints depend on the
particular \zpr \ couplings, but in typical models one requires
$M_{Z^{\prime}} > (500-800) $ GeV and the $Z-\zpr$ mixing angle
$\alpha_{Z-Z^{\prime}}$ to be smaller than a few $\x 10^{-3}$.
Thus, explaining the $Z-Z^{\prime}$ mass hierarchy is important.
Recently, we proposed a supersymmetric model with a
string-motivated  secluded $U(1)^{\prime}$-breaking sector, where
the squark and slepton spectra can mimic those of the MSSM, the
electroweak symmetry breaking is driven by relatively large $A$
terms, and a large \zpr \ mass can be generated by the VEVs of
additional SM singlet fields that are charged under the
\upr~\cite{ELL}.

On the other hand, very light left-handed neutrinos, which has
been confirmed from the solar and atmospheric neutrino
experiments and KamLAND experiment, is a mystery in nature. 
Possible scenarios~\cite{nurevs} with tiny
left-handed neutrino masses include  extensions of the 
SM at a low energy scale, for example, the Zee
model~\cite{Zee}, in which the left-handed neutrino masses are
generated at loop level; supersymmetric models with lepton number
and R-parity violation~\cite{Drees}, which can include both tree
and loop effects; double or extended ({\it i.e.}, TeV-scale) 
see-saw models~\cite{extended};  or models
including large extra dimensions~\cite{lednu}. Mechanisms involving high energy
scales include the canonical see-saw mechanism, in which heavy
right-handed Majorana neutrinos have masses of the order
$10^{10}-10^{16}$  GeV~\cite{Seesaw}; models involving heavy Higgs
triplets~\cite{GRmodel, triplet1, triplet2}; and models 
in which small neutrino Dirac masses are generated by
high-dimensional operators~\cite{hdo,susyint}.

In this paper, we consider the possibilities for small neutrino
masses in supersymmetric $U(1)'$ models. The $U(1)'$ symmetry affects some of the
above mechanisms which can generate the tiny neutrino masses.
 In particular,  right-handed neutrinos
could not acquire  Majorana masses at a scale much larger than the
$U(1)'$-breaking scale unless they are not charged under $U(1)'$,
 thus forbidding a canonical high-scale see-saw mechanism in many
TeV-scale $U(1)'$ models. Another implication involves the
right-handed neutrinos in models with small neutrino Dirac masses. In the
SM these are harmless cosmologically because they are
essentially sterile (except for negligible Higgs couplings and
mass effects) and are not produced in significant numbers prior to
big bang nucleosynthesis (BBN). However, in $U(1)'$ models the right-handed
neutrinos can be produced by these $Z'$ interactions (unless their
$U(1)'$ charge vanishes), leading to stringent constraints on the
$Z'$ mass~\cite{bbn}. Other, comparable, constraints follow from supernova
cooling~\cite{supernova}. 

We discuss a number of possibilities for neutrino masses in
$U(1)'$ models. In Section 2, we
consider the possibility of small Dirac masses. We assume that
elementary renormalizable neutrino Yukawa couplings are forbidden by
the extra gauge symmetry, other symmetries, or string selection
rules, but that effective neutrino Yukawa couplings are generated
by non-renormalizable terms after certain SM singlet
fields acquire intermediate-scale VEVs. Essentially speaking,
this is a generalization of the Froggatt-Nielsen model~\cite{FN}.
We consider models with two additional $U(1)'\times U(1)''$
gauge symmetries, with $U(1)''$
breaking at an intermediate scale and $U(1)'$ at the TeV scale.
The intermediate-scale $U(1)''$ and the associated high-dimensional
operators can account for small neutrino Dirac masses. It can
occur naturally that after the intermediate-scale $U(1)''$ breaking,
the right-handed neutrinos are neutral under the TeV-scale $U(1)'$ so as
to avoid the BBN and supernova constraints. The existence of two 
extra $U(1)$s  is partly
motivated by  $E_6$ grand unification,
since $E_6$ can be broken down to the SM gauge group
with two additional $U(1)$s. However, we only use the
$SU(3)_C\times SU(2)_L\times U(1)_Y \times U(1)'\times U(1)'' \subset E_6$
quantum number assignments for the ordinary and exotic
particles in  ${\bf 27}$ and ${\bf 27}^*$ of $E_6$
 to construct an example of an anomaly free
model, and we do not consider 
the full $E_6$ model\footnote{The Yukawa relations
for a full $E_6$ theory would lead to rapid proton decay for a low
$U(1)'$ breaking scale.}. We describe how the
symmetry breaking pattern can be realized assuming the Higgs fields from
  ${\bf 27}$ and ${\bf 27}^*$ of
$E_6$. We also give an example of how the small neutrino Dirac masses can be
generated in a model with a TeV-scale secluded  $U(1)'$-breaking
sector as in~\cite{ELL}. 
In that case, however, the decoupling of the
right-handed neutrinos requires the introduction of singlets 
not belonging to simple $E_6$ representations. In these models there are
no allowed couplings that can generate large Majorana masses
for the right-handed neutrinos at the intermediate scale. However, we also
consider a variant case in which 
there are allowed couplings which can generate 
 large effective  Majorana masses for the right-handed neutrinos
 through the intermediate-scale $U(1)''$ breaking, 
leading to a traditional see-saw.

Instead of generating  small neutrino Dirac masses from Yukawa couplings
with doublet Higgs fields and high-dimensional operators, one
can generate  small neutrino Majorana masses through their 
couplings with triplet fields. We propose two models involving a
pair of heavy triplets. The mass for the triplets is about
$10^{14}$  GeV for the first model, and about $10^{8}$  GeV for
the second. After the electroweak symmetry breaking, the triplets
obtain very small VEVs and give a realistic left-handed neutrino
Majorana mass matrix. Equivalently, one can integrate out the heavy triplets and
obtain a low energy neutral Higgs potential that is the same as
that in our previous model~\cite{ELL}, up to negligible
corrections, with the left-handed
neutrino Majorana mass matrix generated by the induced non-renormalizable terms.

Yet another possibility is the double-see-saw mechanism. If the
right-handed neutrinos are charged under $U(1)^{\prime}$, they may
acquire Majorana  masses  at  the $U(1)'$-breaking scale.
We consider a
model with the double-see-saw mechanism, in which the neutrino
masses are suppressed by two powers of the TeV-scale masses. The
neutrino Yukawa couplings can be of order $10^{-3}$, {\it i.e.},
the neutrino Dirac masses are of the order of the muon mass. The
double-see-saw mechanism has been discussed previously for one
family in Ref.~\cite{extended}; here, we generalize it
 to three families in the context of $U(1)'$ models.

We slightly modify the model in Ref.~\cite{ELL} by introducing
three right-handed neutrinos and three SM singlets.
Runaway directions can be avoided by imposing suitable conditions
on the soft terms. The vacuum is the same as in~\cite{ELL}, so the
previous discussions on the $Z-Z'$ mass hierarchy and the particle
spectrum still hold. The active neutrino
mass matrix is $M_D (M_V^{-1})^T M_B M_V^{-1} M_D^T$, where $M_D$
is the $3\times 3$ Dirac mass matrix, and $M_V$ and $M_B$ are
$3\times 3$ matrices defined in Section 4. Because the typical
mass scale for $M_V$ is TeV, the active neutrinos may have
realistic masses and mixings if the typical mass scales for $M_B$
and $M_D$ are about 0.1  GeV. We show that normal, inverted and
degenerate textures can be achieved from reasonable assumptions
about $M_D$, $M_V$, and $M_B$.

This paper is organized as follows: in Section 2 we consider the
 supersymmetric $U(1)'\times U(1)''$ models with $U(1)''$ breaking at the
intermediate scale to generate small neutrino Dirac masses. We discuss two
models with a pair of heavy triplets in Section 3. In Section 4,
we consider the supersymmetric 
$U(1)^{\prime}$ model with the double-see-saw mechanism. Our
discussions and conclusions are given in Section 5. We discuss
the runaway directions for the double-see-saw model in Appendix A.

\section{Generating Neutrino Masses from \\
Intermediate-Scale $U(1)''$ Breaking}

We first consider the possibility of generating  small neutrino Dirac
 masses in a $U(1)'\times U(1)''$ models where the $U(1)'$ and $U(1)''$
are broken at the TeV scale and intermediate scale, respectively.
 In this case, we must consider
the constraints from BBN~\cite{bbn}. If the 
right-handed neutrinos are charged under the $U(1)'$, they will couple
to other particles  through the exchange of $U(1)'$ gauge
boson. They must decouple well before the BBN epoch so as to avoid
the BBN constraints from the predicted $^4$He abundance. Either
the $U(1)'$ must be broken at a high scale, typically above 5 TeV, or the
right-handed neutrinos are neutral under the $U(1)'$. Complementary
constraints follow from supernova cooling~\cite{supernova}.
Here we show
that the $N^c_i$ decouplings can occur naturally in certain $U(1)'
\times U(1)''$ models~\cite{sbtalk}.

We consider a model with the gauge group $G_{SM} \times U(1)' \times
U(1)''$, where $G_{SM}$ is the SM gauge group. The
$U(1)''$ is broken at an intermediate scale around $10^{10}$ GeV, the
right-handed neutrinos are left neutral under the $U(1)'$, the
small neutrino Dirac masses are due to high-dimensional operators
associated with the intermediate scale, and the $ U(1)' \times
U(1)''$ symmetry forbids both elementary Majorana masses for the
right-handed neutrinos 
and also renormalizable-level interactions that could generate their
large effective Majorana masses at the intermediate scale.

The $U(1)''$ can be broken at the intermediate scale if it is
associated with a potential which is F- and D- flat at tree level.
For example, if we introduce one pair of vector-like SM
 singlets $S_1$ and $S^*_1$,
 the F-flatness implies a tree-level
potential
\begin{eqnarray}
V(S_1,S^*_1) &=& m_{S_1}^2 |S_1|^2 + 
m_{S^*_1}^2 |S^*_1|^2+{g'^2 Q''^2 \over
2}(|S_1|^2-|S^*_1|^2)^2~,~\,
\end{eqnarray}
where $g'$ is the $U(1)''$ gauge coupling constant and $Q''$
is the $U(1)''$ charge for $S_1$.

For $m_{S_1}^2 + m_{S^*_1}^2 <0$,
 there will be a runaway direction along the
D-flat direction $|\langle S_1 \rangle|=|\langle S^*_1 \rangle|$. 
However, the potential will be
stabilized by loop corrections or high-dimensional
operators, so that the $S_1$ and $S^*_1$
 will obtain intermediate-scale
VEVs. Neutrino Dirac masses could be generated  by high-dimensional
operators, such as
\begin{eqnarray}
W \sim H_2 L_i N_j^c \left({S \over M_{Pl}}\right)^{P_D}~,~\,
\end{eqnarray}
where $L_i$ and $N_j^c$ are the superfields respectively 
corresponding to the
lepton doublets and right-handed neutrinos,
and $M_{Pl}\sim10^{19}$ GeV is the Planck scale.  The $S$ field can be
$S_1$ or $S^*_1$ or any combinations that are allowed by gauge
invariance and other symmetries of the four-dimensional theory,
and by string selection rules. It is reasonable that in some
models neutrino mass terms may occur in higher order than those
for the quarks and charged leptons, leading to naturally small neutrino 
Dirac masses. Choosing proper $S$ field VEVs and powers
$P_D$, one can obtain a realistic neutrino mass spectrum. However,
without a more detailed construction, there is no predictive power
for the type of neutrino hierarchy and the mixing angles.

As an example, we consider how this mechanism can be realized
using the $U(1)' \times U(1)''$ charges 
associated with the ${\bf 27}$ representation of
the $E_6$ gauge group. We show that for the appropriate signs of 
soft supersymmetry breaking parameters, small neutrino Dirac masses can be
generated, with the $N^c_i$ naturally decoupling from the TeV-scale
 $U(1)'$, satisfying the BBN and supernova constraints. We also consider how
to incorporate small neutrino Dirac masses in the secluded $U(1)'$
model. In this case, the decoupling of the $N^c_i$ from the
TeV-scale $U(1)'$ does not occur if all the particles arise from
the simple $E_6$ representations,
 but could in a more general context. We also study
the possibility of generating large Majorana masses for right-handed
 neutrinos in the context of intermediate-scale $U(1)''$
breaking.

\subsection{Neutrino Masses in Models with $E_6$ Particle Content}

Models with the gauge group $G_{SM} \times U(1)' \times U(1)'' $
may appear in grand unification theory with the $E_6$ gauge group,
since $E_6$ can be broken down to the SM  through
\begin{eqnarray}
E_6 \rightarrow SO(10) \times U(1)_{\psi} \rightarrow SU(5) \times
U(1)_{\chi} \times U(1)_{\psi}~.~\,
\end{eqnarray}
The $U(1)_{\chi}$ and $U(1)_{\psi}$ charges for the 
particles in the ${\bf 27}$ representation are
given in Table \ref{E6charge}. The representations of $E_6$ are
automatically anomaly free, so it is an example of a consistent model with
additional $U(1)$s. In a full $E_6$ grand unified theory, the two
extra $U(1)$s would have to be broken at the GUT scale, because
otherwise they would prevent the exotic $D$-quark partners and the
Higgs doublets in the ${\bf 27}$s from acquiring large masses, and the
$D$-scalars could mediate rapid proton decay. Nevertheless, it is
convenient to use the $G_{SM} \times U(1)_{\chi} \times U(1)_{\psi} \subset E_6$
 quantum number assignments for the ordinary and exotic
particles in  ${\bf 27}$ and ${\bf 27}^*$ to construct
 an example of an anomaly free $U(1)' \times
U(1)''$ model,  even
though the rest of $E_6$ structure, such as the Yukawa
relations, is violated. This is typical in string constructions~\cite{srtringval},
which often do not respect the $E_6$ Yukawa relations (that would
be responsible for proton decay), or which may lead to more
complicated $U(1)'$ charge assignments and exotic structure.

\begin{table}[t]
\caption{Decomposition of the $E_6$ fundamental representation
${\bf 27}$ under $SO(10)$, $SU(5)$, and $U(1)$s for the
particles in the ${\bf 27}$. $U(1)_{\chi}$ and $U(1)_{\psi}$ are
orthogonal to each other with $Q_{\chi}=0$ for $S_L$.
$Q_1$ and $Q_2$ are respectively the particle charges under the 
$U(1)_1$ and $U(1)_2$ which are orthogonal with $Q_1=0$ for $\bar{N}$.
 $Q$ is the particle charge under the $U(1)'$ in 
an anomaly free supersymmetric $U(1)'$ model with a secluded
$U(1)'$-breaking sector~\cite{BG}. \label{E6charge}}
\begin{center}
\begin{tabular}{|c| c| c| c| c| c| c|}
\hline $SO(10)$ & $SU(5)$ & $2 \sqrt{10} Q_{\chi}$ & $2 \sqrt{6}
Q_{\psi}$ & $2 \sqrt{15} Q $ & $2 \sqrt{6} Q_2$ & $2 \sqrt{10} Q_1$\\
\hline
16   &   $10~ (u,d,{\bar u}, {\bar e} )$ & $-$1 & 1  & $-{1/2}$ & 1 &1 \\
            &   ${\bar 5}~ ( \bar d, \nu ,e)$  & 3  & 1  & 4  & $-$2 & 2  \\
            &   $1 \bar N$             & $-$5 & 1  & $-$5  & 4 & 0      \\
\hline
       10   &   $5~(D,H^{\prime}_u)$    & 2  & $-$2 & 1  & $-$2 & $-$2     \\
            &   ${\bar 5} ~(\bar D, H^{\prime}_d)$ & $-$2 &$-$2 & $-{7/2}$ & 1 & $-$3\\
\hline
       1    &   $1~ S_L$                  &  0 & 4 & $5/2$ & 1 & 5 \\
\hline
\end{tabular}
\end{center}
\end{table}
We first give an example of a $U(1)' \times U(1)'' $
 model in which $U(1)''$ is broken at
the intermediate scale while the $U(1)'$, which is broken at the TeV
scale, decouples from the right-handed neutrinos. There are two
SM singlets in the ${\bf 27}$, $\bar N$, which we will
identify as the right-handed neutrinos
$N^c_i$, and $S_L$. There
is only one linear combination of $U(1)_{\chi}$ and $U(1)_{\psi}$
in which the $N^c_i$ fields are neutral, that is the $U(1)_1$ shown in
Table \ref{E6charge}\footnote{The $U(1)_1$ charge assignment has been found 
in Ref.~\cite{Ma:1995xk} from  different motivations, {\it i.e.},
to explain the tiny active neutrino masses via 
the high-scale see-saw mechanism or 
allow for the possibility of leptogenesis. However,
leptogenesis is not required since
electroweak baryogenesis in $U(1)'$ models
is a much more viable possibility
than in the MSSM~\cite{BG}.}. $U(1)_2$ is the other linear combination,
which is orthogonal to $U(1)_1$. Here, to avoid confusions,
 we consider the $U(1)_1$ and
$U(1)_2$ as the $U(1)'$ and $U(1)''$, respectively.
We can naturally break the $U(1)_2$ at a high scale by 
giving large VEVs to the scalar components (${\tilde \nu}_R^*$ and
${\tilde \nu}_R$) of
a pair of the vector-like superfields $\nu_R^*$ and $\nu_R$
whose quantum numbers are the same as those of the $N^c_i$ and its
Hermitian conjugate\footnote{The scalar components ${\tilde N}^c_i$
of the right-handed neutrino superfields  $N^c_i$
 should not acquire large VEVs to avoid large lepton-Higgsino
mixings. We therefore assume that the  scalars ${\tilde N}^c_i$
 have positive soft mass-squares,
while the ${\tilde \nu}_R$ and ${\tilde \nu}_R^*$ 
can acquire large VEVs.}.
The F- and D- flatness can be preserved and the 
 $U(1)_1$ unbroken until the TeV scale.
To have D-flat directions, we introduce two pairs of the
vector-like fields $(\nu_R,\nu_R^*)$ and $(S_L, S_L^*)$ from the singlets
of $({\bf 27}, {\bf 27}^*)$  in addition to the SM fermions from
three complete ${\bf 27}$-plets.
Then, the D-term potential is
\begin{eqnarray}
V_{\chi}+V_{\psi} &=& {g'^2 \over 2} \left[{5 \over 2
\sqrt{10}}(|{\tilde \nu}_R|^2-|{\tilde \nu}_R^*|^2)\right]^2 \\&& 
+ {g'^2 \over 2} \left[{1
\over \sqrt{24}}(-|{\tilde \nu}_R|^2+|{\tilde \nu}_R^*|^2-4|S_L|^2
+4|S_L^*|^2)\right]^2~,~\,
\end{eqnarray}
where a sum over each type of scalar is implied, and we have
assumed equal gauge couplings for simplicity. The potential is
clearly D-flat for 
$|\langle {\tilde \nu}_R \rangle|^2=
|\langle {\tilde \nu}_R^* \rangle|^2\equiv |\langle {\tilde \nu} \rangle|^2$ and
$|\langle S_L \rangle|^2 = 
|\langle S_L^* \rangle|^2\equiv |\langle S \rangle|^2$.
We assume that the potential is
also F-flat along this direction. The potential along the flat
direction is then
\begin{eqnarray}
V &=& m_{{\tilde \nu}}^2 |{\tilde \nu}^2|+m_S^2|S^2| ~,~\,
\label{runaway}
\end{eqnarray}
where $m_{{\tilde \nu}}^2$ and $m_S^2$ are respectively the sum of the mass
squares of the ${\tilde \nu}_R$ and ${\tilde \nu}_R^*$, and that of the $S_L$ and
$S_L^*$, which we assume are typical soft-supersymmetry breaking
scale.  For $m_S^2 > 0$ and $m_{{\tilde \nu}}^2<0$, the
breaking will occur along the D-flat direction for
$|\langle {\tilde \nu}_R \rangle|=|\langle {\tilde \nu}_R^* \rangle|$ very large, 
with the  potential ultimately
stabilized by loop corrections or high-dimensional operators~\cite{hdo}.
However, since $m_S^2>0$, $S_L$ and $S_L^*$ will acquire (usually
different) TeV-scale VEVs not associated with the
flat direction.

For arbitrary VEVs, the mass terms for extra gauge bosons are
\begin{eqnarray}
L &=& g'^2\left(-{5 \over 2 \sqrt{10}} Z_{\chi}+{1 \over \sqrt{24}}
Z_{\psi}\right)^2(|{\tilde \nu}_R|^2+|{\tilde \nu}_R^*|^2) 
\nonumber\\&&
+g'^2\left({4\over
\sqrt{24}}Z_{\chi}\right)^2(|S_L|^2+|S_L^*|^2)~.~\,
\end{eqnarray}
For the breaking pattern described above, this will imply that
$\sqrt{\frac{2}{3}}Z_2=-{5\over 2 \sqrt{10}} Z_{\chi}+{1 \over \sqrt{24}} Z_{\psi}$
will acquire a superheavy mass, while the orthogonal combination
$\sqrt{\frac{2}{3}}Z_1={1\over \sqrt{24}}Z_{\chi}+{5\over 2 \sqrt{10}} Z_{\psi}$
will remain at the TeV scale. $Z_1$ decouples from $N^c_i$ and
therefore evades the nucleosynthesis and supernova constraints.

As an example of a high-dimensional operator
to stabilize the potential in Eq. (\ref{runaway}), let us consider
\begin{eqnarray}
W \sim c\frac{(\nu_R \nu_R^*)^2}{M_{Pl}} ~.~\,
\label{hdoop}
\end{eqnarray}
The $\nu_R$ and $\nu_R^*$
fields will obtain  VEVs around $10^{10}/\sqrt{c}$ to $10^{11}/\sqrt{c}$ GeV in this
case\footnote{Alternatively, such an intermediate scale could be generated
by loop corrections to the effective potential, which would render the
running $m_{{\tilde \nu}}^2$ positive at the intermediate scale.}. 
The small neutrino Dirac mass terms may be generated through
\begin{eqnarray}
W \sim H_2 L_i N^c_j \frac{\nu_R \nu_R^*}{M_{Pl}^2}~,~\,
\end{eqnarray}
which is typically of order $10^{-6}/c$
to $10^{-5}/c$ eV. A small $c\sim 10^{-3}-10^{-4}$ would yield appropriate
neutrino masses. Such a value for $c$
could be generated if the operator in Eq. (\ref{hdoop})
was itself due to a high-dimensional operator involving additional fields with
VEVs close to $M_{Pl}$, {\it e.g.}, 
associated with an anomalous $U(1)'$~\cite{anomalous}.

\subsection{Small Neutrino Dirac Masses in 
a $U(1)'$ Model with \\ a Secluded Sector}
The TeV-scale $U(1)'$ could appear in a model with a secluded $U(1)'$-breaking sector
as in~\cite{ELL}. This model can solve the supersymmetric $\mu$
problem, contribute to electroweak baryogenesis, and yield a
$Z-Z'$ hierarchy and small mixing angle. In this
subsection, we show how to extend the $U(1)' \times U(1)''$ model discussed
above to incorporate a secluded sector.

The superpotential for the Higgs sector in the secluded $U(1)'$
model is
\begin{eqnarray}
W_{H} &=& h S H_1 H_2 + \lambda S_1 S_2 S_3 ~,~\,
\end{eqnarray}
where the Yukawa coupling $h$ is
associated with the effective $\mu$ term and the potential has a runaway
direction for $\lambda \rightarrow 0$; the  $S$ and $S_i$ fields are SM
singlets, with $U(1)'$ charge assignments
\begin{eqnarray}
{Q_S=-Q_{S_1} =-Q_{S_2} ={1\over 2} Q_{S_3} ~,~
Q_{H_1}+Q_{H_2}+Q_S=0 ~.~\,} \label{qcharge}
\end{eqnarray}
For a sufficiently small value of $\lambda$, the \zpr \ mass can
be arbitrarily large. For example, if $ h\sim 10 \lambda$, one can
generate a $Z-Z'$ mass hierarchy in which  the $Z'$ mass is of
order 1 TeV~\cite{ELL}.

A $U(1)'$ model with a secluded sector using the
$E_6$ particle contents and charge assignments was  constructed
 in~\cite{BG}. In that
model, it was assumed that the four SM singlets $S$, $S_1$, $S_2$,
$S_3$ are the $S_L$, $S^*_L$, $S^*_L$ and ${\bar N}^*$,
respectively in two pairs of ${\bf 27}$ and ${\bf 27}^*$. Choosing a
special combination of $U(1)_{\chi}$ and $U(1)_{\psi}$ (see the $Q$
charge assignments in Table~\ref{E6charge}), the charge relations
in Eq. (\ref{qcharge}) are satisfied. However, in that model, the 
right-handed neutrinos $N^c_i$ are charged under the low energy $U(1)'$, and
will be constrained by the BBN and supernova data if the neutrinos
have small Dirac masses.

We can instead consider charge assignments such that the right-handed
neutrinos $N^c_i$ will be neutral under the TeV-scale $U(1)'$, {\it i.e.}, the
charge assignments of $Q_1$ and $Q_2$ in Table \ref{E6charge}. As
discussed in the last subsection, the scalar components of the
superfields $\nu_R^*$ and $\nu_R$ 
with the same quantum 
numbers as those of $N^c_i$ and its Hermitian conjugate
 will obtain intermediate-scale
VEVs and break the $U(1)_2$ and leave a TeV-scale  $U(1)_1$. To
incorporate the secluded $U(1)'$ model, one must introduce
SM singlets that satisfy the $U(1)'$ charge relations in Eq. (\ref{qcharge}). 
This is not possible for $S$ or $S_i$ fields belonging to the $({\bf 27}, {\bf 27}^*)$ 
or other low-dimensional
representations of $E_6$. However, recalling that we are using $E_6$ only
as an example of an anomaly-free construction, it is not unreasonable
to consider the possibility of charge assignments for SM singlets that do not
correspond to $E_6$, as long as they are vector-like pairs so as to avoid
anomalies. For example, we can assume the SM singlets $S$,
$S_1$, $S_2$ are the $S_L$, $S_L^*$, $S_L^*$  respectively in
two pairs of ${\bf 27}$ and ${\bf 27}^*$, and also introduce one pair of 
 vector-like fields
$S_3$ and $S_3^*$ with $U(1)'$ charge $Q_{S_3}=-Q_{S_3}^*=2Q_{S_L}$. In this
way, we can generate  small neutrino Dirac
masses from the intermediate-scale $U(1)''$ breaking
in the secluded $U(1)'$-breaking model.

\subsection{Large Majorana Masses 
for Right-Handed Neutrinos}
The above discussions in the $U(1)' \times U(1)'' $ models
concentrated on generating small neutrino Dirac masses
from the intermediate-scale $U(1)''$ breaking. 
One  can also generate the large Majorana
masses for the right-handed neutrinos, to yield the ordinary see-saw
mechanism.

Let us consider three right-handed neutrinos $N^c_i$ and 
one pair of vector-like fields $S$ and $S^*$, with charges,
\begin{eqnarray}
Q'_{N^c_i}=Q'_{S}=0 ~,~
Q''_{N^c_i}~=~-{1 \over 2} Q''_S~,~\,
\end{eqnarray}
where $Q'$ and $Q''$ are the particle charges under
the TeV-scale $U(1)'$ and intermediate-scale $U(1)''$, respectively.
 Then, the superpotential is
\begin{eqnarray}
W \sim {1 \over M_{Pl}^{2k-3}}(S S^*)^k + S N^c_i N^c_j ~.~\,
\end{eqnarray}
We also introduce the soft supersymmetry breaking terms
\begin{eqnarray}
V \sim m_{{\tilde N}_i^c}^2 |{\tilde N}_i^c|^{2} + m_{S}^2
|S|^2+m_{S^*}^2 |S^*|^2~,~
\end{eqnarray}
where ${\tilde N}_i^c$ is the scalar component of
the superfield $N_i^c$.
If we assume $m_{{\tilde N}_i^c}^2 > 0$ while
$m_{S}^2+m_{S^*}^2<0$, the VEVs of the $S$ and $S^*$ fields
 will be driven to
non-zero values, while those of the ${\tilde N}_i^c$ fields will 
be zero. The
D-flat direction will ensure $\langle S \rangle=\langle S^* \rangle$. 
The potential will be
stabilized by the high-dimensional operators, which determines
$\langle S \rangle \sim (m_S M_{Pl}^{2k-3})^{1 \over 2k-2 }$. Taking,
for example, $k=3$ will yield $\langle S \rangle \sim 10^{14}$ GeV.
These VEVs will give  Majorana masses to the $N^c_i$
fields of the same order, allowing an ordinary see-saw mechanism.

\section{Higgs Triplet Models \label{triplet}}

A number of authors have considered models in which small neutrino
Majorana masses can be generated by coupling two lepton doublets
to an $SU(2)_L$-triplet $T$ with weak hypercharge $Y=1$. Early
versions of the triplet models~\cite{GRmodel} assumed spontaneous lepton
number violation. These are excluded by the invisible $Z$ width,
which would be increased equivalent to two extra neutrino species
by $Z$ decaying into the Goldstone boson (Majoron) and a light
scalar. However, more recent scenarios~\cite{triplet1} avoid this difficulty by
coupling  $T$ to the Higgs doublets as well, which breaks lepton
number explicitly. These couplings ensure that $T^0$ acquires a
tiny VEV if $T$ is given a very large mass, or equivalently lead
to the suppressed high-dimensional operators if $T$ is integrated
out. Such models are sometimes referred as the Type II see-saw mechanism.
Supersymmetric versions have been constructed~\cite{triplet2}, and there are
special constraints when this mechanism is embedded in string
constructions, as discussed in~\cite{tripletstring}. Here, we show that the
Type II see-saw mechanism can be applied in the supersymmetric $U(1)'$ models.

We consider the supersymmetric $SU(3)_C \times SU(2)_L \times
U(1)_Y \times U(1)^{\prime}$ models with a pair of very heavy
triplets $T$ and $\bar T$, which can have very small VEVs for the
charge zero components ($T^0$ and $\bar T^0$) after electroweak
symmetry breaking, and give the needed neutrino Majorana masses and
mixings. The quantum numbers for $T$ and $\bar T$ are $(1, 3, 1)$
and $(1, 3, -1)$ under the $SU(3)_C \times SU(2)_L \times U(1)_Y$
gauge symmetry. To be concrete,
 we integrate out the heavy triplets $T$ and $\bar T$, and find that
the low energy neutral Higgs potential is almost the same as that
 in Ref.~\cite{ELL}. Moreover, with suitable Yukawa couplings for
the lepton doublets and triplets, a realistic neutrino Majorana
mass matrix can be generated by non-renormalizable terms. For
simplicity, we only consider the neutral Higgs potential and the
Yukawa couplings for the left-handed neutrino Majorana masses.

\subsection{Model I}

In model I, the $U(1)'$ charges for the Higgs fields, triplets, and
lepton doublets are
\begin{eqnarray}
Q_{H_1} + Q_{H_2} + Q_S = 0 ~,~ Q_L\equiv Q_{L_i}=- Q_{H_2}~,~\,
\end{eqnarray}
\begin{eqnarray}
Q_T=-Q_{\bar T}= 2 Q_{H_2}~.~\,
\end{eqnarray}
We choose the superpotential
\begin{eqnarray}
W &=& h S H_1 H_2  + \lambda_u H_2 \bar T H_2
 + y_{ij} L_i T L_j + m T \bar T ~,~\,
\end{eqnarray}
where $m$ is the mass for $T$ and $\bar T$, which is about
$10^{14}$  GeV. We do not need to introduce
right-handed neutrinos.
Even if there are right-handed neutrinos,
the Yukawa terms $L_i H_2 N_j^c$  are assumed to be
forbidden by the $U(1)'$, or the other symmetries or the underlying
string constructions.
The $F$-term neutral scalar potential is
\begin{eqnarray}
V_F &=& h^2 |H_1^0 H_2^0|^2 + h^2 |S H_2^0|^2 +|h S H_1^0 + 2
\lambda_u {\bar T}^0 H_2^0|^2 
\nonumber\\&& +|\lambda_u H_2^0 H_2^0 + m T^0|^2 + m^2 |{\bar
T}^0|^2~,~\,
\end{eqnarray}
and the $D$-term potential is
\begin{eqnarray}
V_D &=& {{G^2}\over 8} \left(|H_2^0|^2 - |H_1^0|^2 + 2 |T^0|^2 - 2
|{\bar T}^0|^2\right)^2 +{1\over 2} g_{Z'}^2\left(Q_S |S|^2 +
Q_{H_1} |H_1^0|^2 \right.\nonumber\\&&\left. + Q_{H_2} |H_2^0|^2 
 + 2 Q_{H_2} |T^0|^2 -2 Q_{H_2} |{\bar
T}^0|^2 \right)^2 ~,~\,
\end{eqnarray}
where $G^2=g_1^{2} +g_2^2$; $g_1, g_2$,  and $g_{Z'}$ are the
coupling constants for $U(1)_Y, ~SU(2)_L$ and $U(1)^{\prime}$; and
$Q_{\phi}$ is the $U(1)^{\prime}$ charge of the field $\phi$.

We also consider the Yukawa coupling for the neutrinos
\begin{eqnarray}
{\cal L}_{\rm Yukawa} &=&
 - {1\over 2}  y'_{ij} \nu_i T^0 \nu_j + {\rm H.C.}
~,~
\end{eqnarray}
where $\nu_i$ are the left-handed neutrinos, and
$y'_{ij} = y_{ij} (1+\delta_{ij})$ in which $\delta_{ij}$ is equal
to 1 or 0 for $i=j$ or $i\not=j$, respectively.

After electroweak symmetry breaking, {\it i.e.}, $H_1^0 \not= 0$
and $H_2^0 \not=0$, the $F$-terms for $H_2^0$, $T^0$ and ${\bar
T}^0$ cannot be zero simultaneously. The $T^0$ and ${\bar T}^0$
will acquire very small VEVs
\begin{eqnarray}
\langle T^0 \rangle \simeq - {{\lambda_u}\over m}  \langle H_2^0
\rangle
 \langle H_2^0 \rangle
 ~,~  \langle {\bar T}^0 \rangle \simeq - {{2 h \lambda_u^*}\over {m^2}}
\langle S \rangle \langle H_1^0 \rangle \langle H_2^{0*}
\rangle~.~ \, \label{vevttI}
\end{eqnarray}
There are no experimental constraints on the VEVs of $T^0$ and
$\bar T^0$ in this range ({\it i.e.}, much smaller than the electroweak scale). 
The left-handed neutrino Majorana mass
terms are given by
\begin{eqnarray}
{\cal L}_{\rm Yukawa} &=&
 {{\lambda_u}\over {2m}} y'_{ij} \langle H_2^0 \rangle
 \langle H_2^0 \rangle \nu_i \nu_j
+{\rm H. C.}~.~
\end{eqnarray}

Alternatively, we can integrate out the $T^0$ and ${\bar T^0}$
because they are heavy. Their equations of motion are
\begin{eqnarray}
&&m \lambda_u H_2^0 H_2^0 + {1\over 2} y'_{ij}  \nu_i  \nu_j +(m^2
+ 4 \Delta_{EW} + 4 \Delta_{Z'} Q_{H_2})T^0 \nonumber\\&& +(G^2 +
4 g_{Z'}^2 Q_{H_2}^2) |T^0|^2 T^0 =0~,~
\end{eqnarray}
\begin{eqnarray}
&&2 \lambda_u^* h S H_1^0 H_2^{0*} + (m^2 + 4 \lambda_u^2
|H_2^0|^2 - 4 \Delta_{EW} - 4 \Delta_{Z'} Q_{H_2}) {\bar T^0}
\nonumber\\&& + (G^2 + 4 g_{Z'}^2 Q_{H_2}^2) |{\bar T^0}|^2 {\bar
T^0} =0~,~
\end{eqnarray}
where
\begin{eqnarray}
\Delta_{EW} = {{G^2}\over 8} (|H_2^0|^2- |H_1^0|^2)~,~
\end{eqnarray}
\begin{eqnarray}
\Delta_{Z'} = {1\over 2} g_{Z'}^2 \left(Q_S |S|^2 + Q_{H_1}
|H_1^0|^2 + Q_{H_2} |H_2^0|^2 + \sum_{i=1}^3 Q_{S_i}
|S_i|^2\right)~.~
\end{eqnarray}

The terms proportional to $|T^0|^3$ and $|{\bar T^0}|^3$ are very
small due to the large $m$. Thus,
\begin{eqnarray}
T^0 \simeq -{{m \lambda_u H_2^0 H_2^0 + {1\over 2} y'_{ij}  \nu_i
\nu_j} \over\displaystyle {m^2 + 4 \Delta_{EW} + 4 \Delta_{Z'}
Q_{H_2}}} \sim - {{\lambda_u H_2^0 H_2^0} \over\displaystyle m}~,~
\end{eqnarray}
and the resulting non-renormalizable
 neutrino mass terms  are
\begin{eqnarray}
{\cal L}_{\rm Yukawa} &=& {1\over 2} {{ \lambda_u y'_{ij} \nu_i
\nu_j H_2^0 H_2^0}\over\displaystyle m}+ {\rm H.C.} ~.~
\end{eqnarray}

The  neutrino  mass ($m_{\nu}$) scale is about $0.05$ eV,
implying $m\sim 10^{14}$  GeV. With suitable Yukawa couplings
$y_{ij}$,  one can obtain a realistic left-handed neutrino
Majorana mass matrix. Of course, the $U(1)'$ symmetry does not by
itself constrain the form of $y_{ij}$ or lead to a prediction for
the form of the mass hierarchy and mixings. The low energy neutral
Higgs potential is just that in Ref.~\cite{ELL} up to negligible
corrections of order $(M_Z/m)^2 \sim 10^{-24}$.

\subsection{Model II}

In model II, the $U(1)'$ charges for the Higgs fields, triplets
and lepton doublets are
\begin{eqnarray}
Q_{H_1} + Q_{H_2} + Q_S =0 ~,~  Q_{L_i}=Q_L=Q_{H_1} ~,~\,
\end{eqnarray}
\begin{eqnarray}
Q_T=-Q_{\bar T}= - 2 Q_{H_1}~,~\,
\end{eqnarray}
and the superpotential is
\begin{eqnarray}
W &=& h S H_1 H_2  + \lambda_d H_1 T H_1
 + y_{ij} L_i T L_j + m T \bar T ~,~\,
\end{eqnarray}
where $m$ is the mass for $T$ and $\bar T$, around $10^{8}$  GeV.

Similar to the last subsection, after electroweak symmetry
breaking, {\it i.e.}, $H_1^0 \not= 0$ and $H_2^0 \not=0$, the
$F$-terms for $H_1^0$, $T^0$ and ${\bar T}^0$ cannot be zero
simultaneously, and the $T^0$ and ${\bar T}^0$ will have very
small VEVs
\begin{eqnarray}
\langle T^0 \rangle \simeq - {{2 \lambda_d^* h \langle S \rangle
\langle H_2^0 \rangle \langle H_1^{0*} \rangle}\over {m^2}}
 ~,~ \langle {\bar T}^0 \rangle \simeq - {{\lambda_d}\over m}
\langle H_1^0 \rangle \langle H_1^0 \rangle
  ~.~ \,
\label{vevttII}
\end{eqnarray}
The left-handed neutrino Majorana masses are given by
\begin{eqnarray}
{\cal L}_{\rm Yukawa} &=&
 {{1}\over {m^2}} y'_{ij} \lambda_d^* h \langle S \rangle
\langle H_2^0 \rangle \langle H_1^{0*} \rangle \nu_i  \nu_j +{\rm
H. C.}~,~
\end{eqnarray}
which can also be obtained by integrating out $T^0$ and ${\bar
T}^0$.

$m_{\nu} \sim 0.05$ eV can be obtained for $m\sim 10^{8}$
GeV. Suitable Yukawa couplings $y_{ij}$ can yield a
realistic left-handed neutrino Majorana mass matrix, with
negligible small corrections of order $(M_Z/m)^2 \sim 10^{-12}$ to
the low energy Higgs potential.

\section{The Double-See-Saw Mechanism}

Another possibility for small neutrino Majorana masses is the
double or extended see-saw mechanism~\cite{extended}. Typically, the
large scale in such models is only of the order TeV. However, the
light neutrino masses are suppressed by two or more powers of this
scale and sometimes small scales in the numerator. Such
constructions have been suggested, {\it e.g.}, in the context of
superstring model buildings~\cite{stringe}, in which it is difficult to
generate a normal see-saw~\cite{stringseesaw}. They are also a viable possibility
in the $U(1)'$ models, in which the TeV-scale masses may be
associated with the $U(1)'$ breaking scale. In this Section, we
show that the secluded sector model can be extended to include the
double-see-saw mechanism, without introducing unwanted runaway
directions, and that one can obtain the normal and inverted
hierarchies, and the degenerate scenarios, for neutrino masses.

We consider the supersymmetric $SU(3)_C \times SU(2)_L \times
U(1)_Y \times U(1)^{\prime}$ model with 2 Higgs doublets ($H_1$
and $H_2$), 4 Higgs singlets ($S$, $S_1$, $S_2$ and $S_3$), and
three extra singlets ($B_1$, $B_2$ and $B_3$). Assuming
the $U(1)'$ charges satisfy the equations
\begin{eqnarray}
Q_{N_i^c}\equiv Q_{N^c}=-{3\over 2} Q_S ~,~ Q_{B_i}\equiv Q_B=-{1\over 2} Q_S ~,~\,
\end{eqnarray}
\begin{eqnarray}
Q_{L_i}\equiv Q_L=- Q_{H_2}+{3\over 2}
Q_S~,~\,
\end{eqnarray}
as well as these in Eq. (\ref{qcharge}),
we choose the superpotential
\begin{eqnarray}
W &=& h S H_1 H_2 + \lambda S_1 S_2 S_3 +d_{ij} S B_i B_j
 + e_{ij} S_3 N_i^c B_j + y_{ij} H_2 N_i^c L_j  ~,~\,
\end{eqnarray}
where $h$, $\lambda$, $d_{ij}$, $e_{ij}$ and $y_{ij}$ are Yukawa
couplings, and we assume that $d_{ii}=0$, motivated by string
constructions.
The corresponding $F$-term scalar potential is
\begin{eqnarray}
V_F &=& h^2 |S|^2 |H_2|^2 + |h S H_1 + y_{ij} {\tilde N}_i^c {\tilde
L}_j|^2 +|h H_1 H_2 + d_{ij} {\tilde B}_i {\tilde B}_j|^2
\nonumber\\&& +\lambda^2 \left(|S_2|^2 + |S_1|^2\right) |S_3|^2
+|\lambda S_1 S_2 + e_{ij} {\tilde N}_i^c {\tilde B}_j|^2
\nonumber\\&& +\sum_{i=1}^3 |e_{ij} S_3 {\tilde B}_j + y_{ij} H_2
{\tilde L}_j|^2 +\sum_{j=1}^3 |y_{ij} H_2 {\tilde N}_i^c|^2
\nonumber\\&& +\sum_{j=1}^3 |d_{ij} S {\tilde B}_i + e_{ij} S_3
{\tilde N}_i^c|^2 ~,~\,
\end{eqnarray}
where for a supermultiplet $\phi$ which is not a Higgs doublet
($H_1$ or $H_2$) or
singlet  field ($S$ or $S_i$), we denote
 its scalar component as $\tilde \phi$.
The $D$-term scalar potential for the fields that are $SU(3)$
singlets and neutral under $U(1)_Y$ is
\begin{eqnarray}
V_D &=& {{G^2}\over 8} \left(|H_2^0|^2 - |H_1^0|^2 - \sum_{i=1}^3
|{\tilde \nu}_i|^2  \right)^2 \nonumber\\&& +{1\over 2}
g_{Z'}^2\left(Q_S |S|^2 + Q_{H_1} |H_1^0|^2 + Q_{H_2} |H_2^0|^2
\right.\nonumber\\&&\left. + \sum_{i=1}^3 Q_{S_i} |S_i|^2 +
\sum_{i=1}^3 (Q_{N} |{\tilde N}_i^c|^2 + Q_{L} |{\tilde \nu}_i|^2
+Q_B |{\tilde B}_i|^2 )\right)^2 ~.~\,
\end{eqnarray}
In addition, we introduce the supersymmetry breaking soft terms
\begin{eqnarray}
V_{soft} &=& m_{H_1}^2 |H_1|^2 + m_{H_2}^2 |H_2|^2 + m_S^2 |S|^2
\nonumber\\&&
 + \sum_{i=1}^3 \left( m_{S_i}^2 |S_i|^2 +m_{{\tilde N}_i^c}^2 |{\tilde N}_i^c|^{2}
+m_{{\tilde \nu}_i}^2 |{\tilde \nu}_i|^2 + m_{{\tilde B}_i}^2
|{\tilde B}_i|^2\right) \nonumber\\&& -\left(A_h h S H_1 H_2 +
A_{\lambda} \lambda S_1 S_2 S_3 + A_{d_{ij}}  d_{ij} S {\tilde
B}_i {\tilde B}_j
 + A_{e_{ij}}  e_{ij} S_3 N_i^c {\tilde B}_j
\right.\nonumber\\&&\left. + A_{y_{ij}} y_{ij} H_2 N_i^c L_j + {\rm
H. C.}\right) + (m_{S S_1}^2 S S_1 + m_{S S_2}^2 S S_2 + {\rm H.
C.}) ~.~\, \,  \label{vsoft}
\end{eqnarray}
For simplicity, we do not consider the soft mass terms like
$S_1^{\dagger} S_2$ or $ {\tilde N}_i^{c\dagger} {\tilde N}_j^c$ or $
{\tilde B}_i^{\dagger} {\tilde B}_j$, etc.

The runaway directions for the unbounded from below scalar
potential
 are discussed in Appendix A, where
 suitable conditions to avoid them are given.
Because we choose relatively large and positive soft mass-squares
for ${\tilde \nu}_i$, ${\tilde N}_i^c$ and ${\tilde B}_i$ that are
of order 200  GeV or $A_h$, the scalar fields
 ${\tilde \nu}_i$, ${\tilde N}_i^c$ and ${\tilde B}_i$ do not
acquire non-zero VEVs. Thus, the VEVs for the $H_1^0$, $H_2^0$, $S$
and $S_i$ are the same as those in~\cite{ELL}, and the $Z-Z'$ mass
hierarchy and the particle spectrum for charginos, neutralinos and
Higgs particles are unchanged.

In the basis $\{\nu_1, \nu_2, \nu_3, N_1^c, N_2^c, N_3^c, B_1, B_2,
B_3\}$, the neutrino mass matrix is
\begin{eqnarray}
M = \left(\matrix{ 0 & M_D &0 \cr M_D^T & 0 & M_V \cr 0 & M_V^T &
M_B \cr}\right) ~,~ \,
\end{eqnarray}
where
\begin{eqnarray}
\left( M_D \right) = y'_{ij} v_2 ~,~ \left( M_V \right) = e_{ij}
s_3~,~ \left( M_B \right) = d_{ij} s/2~,~ \,
\end{eqnarray}
with $\langle H_2^0 \rangle = v_2$, $\langle S_3 \rangle = s_3 $,
$\langle S \rangle =s$, and the upper index $T$ denotes the
transpose.

Define the matrix $U$ as
\begin{eqnarray}
U = \left(\matrix{ 1 & {{\sqrt 2} \over 2} M_D^*
(M_V^{-1})^{\dagger} & {{\sqrt 2} \over 2} M_D^*
(M_V^{-1})^{\dagger} \cr (M_V^{-1})^T M_B M_V^{-1} M_D^T & {{\sqrt
2} \over 2} & - {{\sqrt 2} \over 2} \cr -M_V^{-1} M_D^T & {{\sqrt
2} \over 2} & {{\sqrt 2} \over 2} \cr}\right) ~.~ \,
\end{eqnarray}
 $U^T M U$ is approximately (up to ${\cal O}(M_V^{-3})$)
block diagonal with an upper $3\times 3$ block which gives 3 very
light active neutrinos, and a $6\times 6$ block which gives 6
heavy SM singlets. The $3\times 3$ matrix for the active neutrinos
is
\begin{eqnarray}
M_{\nu} &=& M_D (M_V^{-1})^T M_B M_V^{-1} M_D^T ~.~ \,
\label{nmass}
\end{eqnarray}
Using our previous numerical results for the vacuum in
Ref.~\cite{ELL}, we have that $v_2 \sim 125$  GeV, $s \sim 187$
GeV and $s_3 \sim 1260$  GeV. Therefore,  realistic active
neutrino masses can be obtained, {\it e.g.},  for $y_{ij}$ and $d_{ij}$ of order
$10^{-3}$, and  $e_{ij}$ of order  $1$.

We consider the real case for simplicity. The discussions for the
complex case are similar. $M_D$ and $M_V$ are general $3\times 3$
mass matrices which have 9 independent parameters, and $M_B$ is a
symmetric matrix without diagonal entries, which has 3 independent
parameters. However, only  $M_D (M_V^{-1})^T$ and $M_B$ enter the
expression for $M_{\nu}$, so, there are $9+3=12$ independent
parameters. Using Mathematica, one can show that $M_{\nu}$ is
equivalent to a general real and symmetric mass matrix for the
active neutrinos, which
 has 6 independent parameters.

It is not hard to find examples which lead to realistic
 neutrino mass matrices.
Here, we consider simple patterns corresponding to a normal
hierarchy, an inverted hierarchy with the same signs
 for the eigenvalues $m_{\nu_1}$ and $m_{\nu_{2}}$,
an inverted hierarchy with opposite signs, and the degenerate
case.

Define the matrices
\begin{eqnarray}
\alpha = \left(\matrix{ 0 & 0 &0 \cr 0 & 1 & 1 \cr 0 & 1 & 1
\cr}\right) ~,~ \beta = \left(\matrix{ 2 & 0 &0 \cr 0 & 1 & -1 \cr
0 & -1 & 1 \cr}\right) ~,~ \gamma = \left(\matrix{ 0 & 1 &-1 \cr 1
& 0 & 0 \cr -1 & 0 & 0 \cr}\right) ~,~ \,
\end{eqnarray}
which correspond to the zeroth order approximations for the
patterns of the  normal hierarchy, inverted hierarchy with same
sign eigenvalues, and inverted hierarchy with opposite
 sign  eigenvalues, respectively.
$\alpha$ and $\beta$ lead to maximal atmospheric neutrino mixing,
with the solar neutrino mixing depending on the subleading terms
(not displayed) and on the charged lepton mixings. $\gamma$ leads
to bimaximal mixings, which can be consistent with the observed
(non-maximal) solar neutrino mixing if there is small
(Cabibbo-like) mixing in the charged lepton sector~\cite{charged}. Define the
mass matrix $M'_{\nu}$ as
\begin{eqnarray}
M'_{\nu}= X \alpha + Y \beta + Z\gamma ~.~ \,
\end{eqnarray}
For simplicity, we consider the scenarios in which $M_D$, $M_V$
and $M_B$ are order unity, {\it i.e.}, the magnitudes of
the entries are ${\cal O}(1)$ or ${\cal O}(0)$, and show that one
can
  produce the above simple patterns and the patterns with degenerate masses.
One can use the freedom in the right hand side of Eq.
(\ref{nmass}) to choose
\begin{eqnarray}
M_V=\left(\matrix{ 1 & 0 &0 \cr 0 & 1 & 0 \cr 0 & 0 & 1
\cr}\right) ~,~ \,
\end{eqnarray}
\begin{eqnarray}
M_B=\left(\matrix{ 0 & 1 & 1 \cr 1 & 0 & -1 \cr 1 & -1 & 0
\cr}\right) ~,~ \,
\end{eqnarray}
\begin{eqnarray}
M_D=\left(\matrix{ {\sqrt {(d-e) f + d (e+f)}} & b & {\sqrt {(d-e)
f + d (e+f)}}  \cr d &  e & f \cr -d & e & -f \cr}\right) ~.~ \,
\end{eqnarray}
Requiring that $M_{\nu}= M'_{\nu}$, we obtain
\begin{eqnarray}
X=(d-f)e ~,~ Y= (d-f) e + 2 d f ~,~ \, \label{xyz1}
\end{eqnarray}
\begin{eqnarray}
Z=b(d-f)+(d+f) {\sqrt {(d-e) f + d (e+f)}}  ~.~ \, \label{xyz2}
\end{eqnarray}
In the following, we give the solutions for five simple patterns:

(1) The normal hierarchy: $X\not=0$ and $Y=Z=0$. A simple solution
to Eqs. (\ref{xyz1}) and (\ref{xyz2}) is that $b=0$ and
$d=e=-f=\sqrt {X/2}$, so the Dirac mass matrix $M_D$ is
\begin{eqnarray}
M_D = {\sqrt {X\over 2}} \left(\matrix{ 0 & 0 & 0 \cr 1  & 1  & -1
\cr -1 & 1 & 1 \cr}\right) ~.~ \,
\end{eqnarray}

(2) The inverted hierarchy with same
 signs for the eigenvalues $m_{\nu_1}$ and $m_{\nu_{2}}$: $Y\not=0$ and $X=Z=0$.
A simple real solution is $b=-3 \sqrt Y$, $d=\sqrt Y$, $e=0$
and $f={\sqrt Y}/2$, implying
\begin{eqnarray}
M_D = {{\sqrt Y} \over 2} \left(\matrix{ 2 & -6 & 2 \cr 2  & 0  &
1 \cr -2 & 0 & -1 \cr}\right) ~.~ \,
\end{eqnarray}
For $M_D$ complex, there is a simple solution in which $b=0$,
$d=-f= i {\sqrt {Y/2}}$ and $e=0$, so,
\begin{eqnarray}
M_D  = {\sqrt {Y\over 2}} \left(\matrix{ \sqrt 2 & 0 & \sqrt 2 \cr
i  & 0  & -i \cr -i & 0 & i \cr}\right) ~.~ \,
\end{eqnarray}

(3) The inverted hierarchy with opposite signs: $Z\not=0$ and
$X=Y=0$. A simple solution is  $b=d={\sqrt Z}$, $e=0$ and $f=0$, thus,
\begin{eqnarray}
M_D = {\sqrt Z} \left(\matrix{ 0 & 1 & 0 \cr 1  & 0  & 0 \cr -1 &
0 & 0 \cr}\right) ~.~ \,
\end{eqnarray}

(4) The degenerate scenario: $X=Y\not=0$ and $Z=0$,
 can be obtained for $b=-d=-e=-{\sqrt X}$ and $f=0$, with
\begin{eqnarray}
M_D = {\sqrt {X}} \left(\matrix{ 1 & -1 & 1 \cr 1  & 1  & 0 \cr -1
& 1 & 0 \cr}\right) ~.~ \,
\end{eqnarray}

(5) The degenerate scenario: $X=Z\not=0$ and $Y=0$, corresponds to
 $b=d=e=-f= {\sqrt {X/2}}$, yielding
\begin{eqnarray}
M_D = {\sqrt {X\over 2}} \left(\matrix{ 0 & 1 & 0 \cr 1  & 1  & -1
\cr -1 & 1 & 1 \cr}\right) ~.~ \,
\end{eqnarray}

\section{Discussions and Conclusions}
\label{discussion} 
In this paper, we considered neutrino masses
in supersymmetric models with an additional TeV-scale $U(1)'$ gauge
symmetry, in which the ordinary see-saw mechanism may not work
unless the right-handed neutrinos have no $U(1)'$ charge. 
We proposed three mechanisms for neutrino masses in
such models. First, in models with the gauge group $G_{SM} \times
U(1)' \times U(1)''$, with the $U(1)''$ breaking at the 
intermediate scale, the neutrinos may obtain small Dirac masses through
high-dimensional operators associated with the intermediate
scale. We illustrated this mechanism in a model with the $E_6$
particle content and charge assignments, and showed that the right-handed
neutrinos could naturally decouple from the TeV-scale
$U(1)'$, thus avoiding cosmological and astrophysical constraints. We also
discussed this mechanism for models with a secluded
$U(1)^{\prime}$-breaking sector (in which the $Z-Z'$ mass hierarchy
can be generated naturally) and an intermediate-scale $U(1)''$. In
this case the right-handed neutrinos are charged under
$U(1)^{\prime}$  unless one goes outside of the $E_6$
framework for the charge assignments of the SM singlets.
 We also considered the possibility
that the large Majorana masses for right-handed neutrinos
 can be generated through the
intermediate-scale $U(1)''$ breaking, leading to an ordinary
see-saw.

In addition, we described two models with pairs of heavy
triplets, with masses around $10^{14}$ GeV and $10^{8}$ GeV, respectively.
After the electroweak symmetry breaking, the triplets
obtain very small VEVs and can give a reasonable left-handed
neutrino Majorana mass matrix. One can instead  integrate out the heavy
triplets and obtain the small left-handed neutrino Majorana masses
from the resulting non-renormalizable operators. The low energy
neutral Higgs potential is the same as that in \cite{ELL} up to
negligible corrections.

We also studied models in which very small neutrino Majorana masses can be
obtained by the double-see-saw mechanism. The neutrino Yukawa
couplings can be of order $10^{-3}$, {\it i.e.}, the neutrino
Dirac masses are comparable to the muon mass. We slightly modified
the model in Ref.~\cite{ELL} by introducing three right-handed
neutrinos and three SM singlets. Runaway diections can
be avoided by imposing suitable conditions on the soft terms. The
vacuum is the same as in~\cite{ELL}, so the $Z-Z'$ mass hierarchy
and the particle spectrum are not modified. The active
neutrino mass matrix is $M_D (M_V^{-1})^T M_B M_V^{-1} M_D^T$,
where the typical mass scale for $M_V$ is TeV (the $U(1)'$-breaking
 scale). The active neutrinos can have realistic masses
and mixings if the typical mass scales for $M_B$ and $M_D$ are
about 0.1  GeV. Specific examples for the form of the neutrino 
Dirac mass matrix
that lead to normal and inverted hierarchies and to the
degenerate scenario are given.

\section*{Acknowledgments}
T. Li would like to thank V. Barger and Hong-Jian He for useful
discussions on neutrino physics. This research was supported in
part by the U.S.~Department of Energy under
 Grant No.~DOE-EY-76-02-3071.
The research of T. Li was also supported by the National Science
Foundation under
 Grant No.~PHY-0070928.

\section*{Appendix A}

We first consider the scalar potential in Section 4 without
neutrino Yukawa couplings, {\it i.e.}, in the limit $y_{ij}=0$.
The condition for 6 heavy SM singlets at the TeV scale
is $\det|e_{ij}| \not=0$. The discussions of the unbounded from
below runaway directions for the scalar potential are standard,
so, we will not give the details. The constraint conditions to
avoid them are
\begin{equation}
m_S^2 + m_{S_1}^2- 2  |m_{SS_1}^2| > 0 ~,~ \, \label{cc1}
\end{equation}
\begin{equation}
m_S^2 + m_{S_2}^2- 2  |m_{SS_2}^2| > 0 ~,~ \, \label{cc2}
\end{equation}
\begin{equation}
\frac{2}{3} m_S^2 + m_{{\tilde N^c}_i}^2 > 0 ~,~ \, \label{cc3}
\end{equation}
\begin{equation}
\frac{3}{2} m_S^2- \frac{3}{2}\mid m_{SS_1}^2 \mid + m_{{\tilde
N^c}_i}^2 > 0 ~,~ \, \label{cc4}
\end{equation}
\begin{equation}
\frac{3}{2} m_S^2- \frac{3}{2}\mid m_{SS_2}^2 \mid + m_{{\tilde
N^c}_i}^2 > 0 ~,~ \, \label{cc5}
\end{equation}
where $i=1, 2, 3$. The constraint condition in Eq. (\ref{cc1}) (or
Eq. (\ref{cc2})) avoids runaway directions in which $\langle S
\rangle$ and $\langle S_1 \rangle$ (or $\langle S_2 \rangle$) go
to infinity while the other fields have finite VEVs.  Eq.
(\ref{cc3}) avoids the runaway directions in which  $\langle S
\rangle$ and $\langle {\tilde N}_i^c \rangle$ (or two or three
$\langle {\tilde N}_i^c \rangle$)  go to infinity. The condition in
Eq. (\ref{cc4}) (or Eq. (\ref{cc5})) avoids the runaway directions
for which $\langle S \rangle$, $\langle S_1 \rangle$ (or $\langle S_2
\rangle$),
 and $\langle {\tilde N}_i^c \rangle$ (or two or three $\langle {\tilde N}_i^c \rangle$)  go to infinity.
We assume that $m_{{\tilde N^c}_i}$ are positive, so Eq. (\ref{cc3})
is satisfied automatically if Eq. (\ref{cc4}) or Eq. (\ref{cc5})
is satisfied.

Now let us include the neutrino Yukawa couplings. For simplicity,
we assume that $A_{d_{ij}}=A_{e_{ij}}=A_{y_{ij}}=0$. The only new
possible runaway directions have $\mid \nu_i\mid \to \infty$, {\it
i.e. }, $\langle {\tilde \nu}_i \rangle$, $\langle S_3 \rangle$,
$\langle {\tilde B}_i \rangle$, $\langle H_1^0 \rangle$ and
$\langle H_2^0 \rangle$ can go to infinity, while the other fields
have finite VEVs. The VEVs for $\langle {\tilde \nu}_i \rangle$,
$\langle S_3 \rangle$, $\langle {\tilde B}_i \rangle$, $\langle
H_1^0 \rangle$  and $\langle H_2^0 \rangle$ must satisfy 
\begin{equation}
e_{ij} \langle S_3 \rangle  \langle {\tilde B}_j \rangle = -
h_{ij} \langle H_2^0 \rangle \langle {\tilde \nu}_j \rangle ~,~\,
\label{cc6}
\end{equation}
\begin{equation}
h \langle H_1^0 \rangle \langle H_2^0 \rangle = -d_{ij} \langle
{\tilde B}_i \rangle \langle {\tilde B}_j \rangle  ~,~ \,
\label{cc7}
\end{equation}
\begin{equation}
\mid \langle H_2^0 \rangle\mid^2=\mid \langle H_1^0
\rangle\mid^2+\sum_{i=1}^3\mid \langle  {\tilde \nu}_i  \rangle
\mid^2 ~,~ \, \label{cc8}
\end{equation}
\begin{eqnarray}
\frac{1}{2} \sum_{i=1}^3 \mid \langle {\tilde B}_i \rangle \mid^2
&=& 2 \mid \langle S_3 \rangle \mid ^2 +Q_{H_1}\mid \langle H_1^0
\rangle \mid^2+Q_{H_2}\mid \langle H_2^0 \rangle \mid^2
\nonumber\\&& +\sum_{i=1}^3  Q_L \mid \langle {\tilde \nu}_i
\rangle \mid^2 ~.~ \, \label{cc9}
\end{eqnarray}
The potential is very complicated, so we will impose strong
conditions to avoid the runaway direction. For the $F$-terms, we only
keep the term $h^2 \mid S\mid^2 \mid H_2\mid^2$ because $F$-terms
give positive contributions to the potentials. Using Eq.s
(\ref{cc6}), (\ref{cc7}), (\ref{cc8}) and (\ref{cc9}), we obtain
\begin{eqnarray}
 V_{{\rm total}} &>&
\left(m_{H_1}^2+m_{H_2}^2+\frac{m_{S_3}^2}{2}-A_h^2\right) \mid H_1^0\mid ^2
+\left(m_{{\tilde B}_i}^2
+\frac{m_{S_3}^2}{4} \right) \mid {\tilde B}_i\mid^2
\nonumber\\&&
 +\left(m_{{\tilde \nu}_i}^2+m_{H_2}^2+\frac{Q_L+Q_{H_2}}{2}
m_{S_3}^2\right) \mid
{\tilde \nu}_i\mid ^2 + ~{\rm constant},~\, 
\end{eqnarray}
where $V_{{\rm total}}= V_F+V_D+V_{soft}$.
Because $|\langle H_1^0 \rangle| <  |\langle H_2^0 \rangle|$ and
we consider the large $A_h$ scenario in which
$m_{H_1}^2+m_{H_2}^2+\frac{m_{S_3}^2}{2}-A_h^2 < 0$, we obtain
\begin{eqnarray}
 V_{{\rm total}} &>&
\left(m_{H_1}^2+m_{H_2}^2+\frac{m_{S_3}^2}{2}-A_h^2\right)\frac{1}{h}\mid
d_{ij} \mid \mid {\tilde B}_i\mid \mid {\tilde B}_j
\mid + \left(m_{{\tilde B}_i}^2+\frac{m_{S_3}^2}{4}\right) \mid {\tilde B}_i\mid^2 
\nonumber\\&&
 +\left(m_{{\tilde \nu}_i}^2+m_{H_2}^2-\frac{Q_L+Q_{H_2}}{2}
m_{S_3}^2\right) \mid
{\tilde \nu}_i\mid ^2  +  ~{\rm constant}.~\,
\end{eqnarray}
To avoid the runaway directions, we require
\begin{equation}
m_{{\tilde \nu}_i}^2+m_{H_2}^2 - \frac{Q_L+Q_{H_2}}{2} m_{S_3}^2 >
0~,~ \,
\end{equation}
\begin{equation}
m_{{\tilde B}_i}^2 + \frac{m_{S_3}^2}{4} >
{{A_h^2-m_{H_1}^2-m_{H_2}^2-\frac{m_{S_3}^2}{2}}
\over\displaystyle {h}}
 ~{\rm max}\{ \mid d_{ij} \mid,
\mid d_{ik} \mid \} ~,~ \,
\end{equation}
where $i\not=j\not=k$ and $i, j, k =1, 2, 3$.

Because  $m_{{\tilde \nu}_i}^2$, $m_{{\tilde N^c}_i}^2$ and
 $m_{{\tilde B}_i}^2$ are relatively large positive soft mass squares of the order
of $A_h^2$, the above conditions are satisfied.


\end{document}